\documentclass[aps,prb,twocolumn,floats,final,superscriptaddress,citeautoscript,longbibliography]{revtex4-2}
\usepackage{amssymb}
\usepackage{graphicx}
\usepackage{amsmath}
\usepackage{amsthm}
\usepackage{braket,physics}
\usepackage[colorlinks=true,linkcolor=blue,citecolor=blue,urlcolor=blue]{hyperref}
\usepackage{titlesec}

\begin{document}

\title{Quantum many-body scars through the lens of correlation matrix}

\author{Zhiyuan Yao}
\email{yaozy@lzu.edu.cn}
\affiliation{Key Laboratory of Quantum Theory and Applications of MoE, Lanzhou Center for Theoretical Physics, and Key Laboratory of Theoretical Physics of Gansu Province, Lanzhou University, Lanzhou, Gansu 730000, China}
\author{Pengfei Zhang}
\affiliation{Department of Physics, Fudan University and State Key Laboratory of Surface Physics, Shanghai, 200438, China}
\affiliation{Shanghai Qi Zhi Institute, AI Tower, Xuhui District, Shanghai 200232, China}
\affiliation{Hefei National Laboratory, Hefei 230088, China}

\date{\today}

\begin{abstract}
Quantum many-body scars (QMBS)---rare eigenstates that evade thermalization---are typically characterized by their low entanglement entropies compared to surrounding thermal eigenstates. However, due to finite-size effects in systems accessible via exact diagonalization, this measure can be ambiguous.
To address this limitation, we propose using the correlation matrix spectrum as an alternative probe to identify QMBS.
In cases of exact QMBS that either have known analytic expressions or are captured by various frameworks of QMBS, we find that the dimensionality of the null space of the correlation matrix---an integer value, and thus immune to finite-size effects---can qualitatively identify QMBS.
Beyond serving as a diagnostic tool, the correlation matrix method enables the manipulation of the QMBS subspace.
For approximate QMBS, such as those in the PXP model, we observe that the correlation matrix spectrum features numerous approximate zero eigenvalues, thereby distinguishing these states.
We demonstrate the effectiveness and utility of this method with several paradigmatic QMBS examples.
\end{abstract}

\maketitle

\section{INTRODUCTION}
With rapid advances in preparing and manipulating quantum states across various artificial quantum platforms, there has been growing interest in understanding how a generic isolated quantum system thermalizes after a quench \cite{Nandkishore_2015mb,Borgonovi_2016qc,DAlessio_2016fq,Mori_2018ta,Abanin_2019cm,Ueda_2020qe}.
The universal thermalization imposes stringent constraints on the eigenstates of generic interacting quantum systems, and a widely accepted framework is the eigenstate thermalization hypothesis (ETH), which
posits that, locally, an eigenstate encodes the information of a thermal ensemble \cite{Deutsch_1991qs,Srednicki_1994ca,Srednicki_1999ta,Dunjko_2012tf,Deutsch_2018et}.
Except for systems that are integrable \cite{Sutherland_Book,Franchini_Book,Rigol_2007ri,Vidmar_2016gg} or with emergent integrability, such as the many-body localized systems \cite{Gornyi_2005ie,Basko_2006mi,Abanin_2019cm},
the validity of ETH has been confirmed in numerous numerical studies \cite{Rigol_2008ta,DAlessio_2016fq}. Moreover, it was believed that the ETH applies in the \emph{strong sense} that it holds for every eigenstate with finite energy density \cite{Kim_2014tw}.

However, the ETH has recently been challenged by persistent oscillations of local observables after a quench in a Rydberg atom chain \cite{Bernien_2017pm}.
Shortly after, it was shown that the ETH in the effective PXP model is valid only in the \emph{weak sense}: it breaks down for a measure-zero subset of eigenstates \cite{Turner_2018we,Turner_2018qs,Ho_2019po,Choi_2019es,Lin_2019eq,Khemani_2019so,Iadecola_2019qm,Bull_2019sc,Turner_2021cp,Surace_2021em,Windt_2022sq,Yao_2022qm,Pan_2022cs,Desaules_2022em,Desaules_2022hs,Omiya_2023qm,Daniel_2023bq,Kerschbaumer_2024qm}.
These special outliers, drawing analogy to single-particle phenomena \cite{Heller_1984bs}, were dubbed quantum many-body scars (QMBS), or simply scar states.
This discovery sparked intense ongoing research, significantly expanding our understanding of QMBS \cite{Serbyn_2021qm,Moudgalya_2022qm,Chandran_2023qm}.
For example, Shiraishi and Mori have constructed a framework to embed a target space spanned by scar states in a nonintegrable model \cite{Shiraishi_2017sc}.
Families of Hamiltonians hosting QMBS with closed analytic expressions and equidistant energy separation have also been constructed from symmetry and quasisymmetry perspectives \cite{Pakrouski_2020mb,Pakrouski_2021gt,Sun_2023ms,Ren_2021qg,Ren_2022ds,ODea_2020ft}.
In parallel, in the spirit of $\eta$-pairing states in the Hubbard model \cite{Yang_1989ep,Yang_1990ss}, the Mark--Lin--Motrunich (MLM) framework  \cite{Mark_2020us} and the equivalent spectrum generating algebra (SGA) formalism \cite{Moudgalya_2020ep}
have provided unified explanations of scar states discovered across different models
\cite{Moudgalya_2018ee,Moudgalya_2018eo,Moudgalya_2020lc,Mark_2020ep,Schecter_2019we,Iadecola_2020qm,Shibata_2020os}. Recently, using the language of commutant algebra, an endeavor to construct ``exhaustive'' Hamiltonians for a set of QMBS was made by S. Moudgalya and O. I. Motrunich \cite{Moudgalya_2024ec}. Moreover, a necessary condition to characterize an exact QMBS as the eigenstate of multiple noncommuting local operators has been proposed \cite{Moudgalya_2024ec}.

Despite all these advances, a simple yet unified framework for QMBS is still lacking \cite{Moudgalya_2022qm}. For example, although the commutant algebra framework unifies several previous formalisms, it is unclear whether the SGA formalism can be captured by the commutant algebra approach or not \cite{Moudgalya_2024ec}.
Additionally, controlling the scar subspace is often challenging.
On the numerical side, the entanglement entropy is commonly used to identify QMBS due to their characteristic low-entanglement nature \cite{Papic_2022we}. However, because of the limited system sizes accessible via exact diagonalization, finite-size effects can make it difficult to reliably distinguish QMBS using this approach.

In this paper, we show that these problems can be well addressed using the correlation matrix spectrum as a tool for characterizing and diagnosing QMBS. For exact QMBS, including those not captured by previous formalisms, we find that the dimensionality of the kernel of the correlation matrix---being an integer and immune to finite-size effects---differs from that of thermal eigenstates. This number thus provides a sharp \textit{qualitative} probe of QMBS, superior to the \textit{quantitative} probe of entanglement entropy. Compared to the necessary condition of exact QMBS in the commutant algebra framework \cite{Moudgalya_2024ec}, our approach motivates a definition of, or a sufficient condition for, exact QMBS. For approximate QMBS, such as those in the PXP model, the distinct feature of numerous approximate zero eigenvalues in the correlation matrix spectrum also allows us to single out QMBS from thermal eigenstates.

This paper is organized as follows. In Sec.~\ref{sec:CorrMat}, we provide a brief introduction to the correlation matrix method and its underlying information perspective on the ETH. Sec.~\ref{sec:exact_QMBS} is dedicated to the correlation matrix study of the generalized AKLT model where we demonstrate that correlation-matrix zeros, $N_{0}$, serve as a sharp probe of QMBS. In addition to uncovering extra series of QMBS in this model, an additional 2-local term that preserves the SGA scar tower has been identified and is related to the recently discovered staggered projection eigenoperator \cite{Rozon_2024bu}.
In Sec.~\ref{sec:embedding}, we discuss the manipulation of QMBS and the inner constraints imposed by the kernels of correlation matrices. In Sec.~\ref{sec:unification}, we show that our correlation matrix approach unifies many previous paradigmatic formalisms, including the SGA formalism. In Sec.~\ref{sec:approximate_QMBS}, we discuss how the spectrum of a correlation matrix can be used to quantitatively distinguish QMBS from their surrounding thermal eigenstates, playing a similar role to entanglement entropy.
In Sec.~\ref{sec:thermal}, we explore the implications of our findings for the structure of thermal eigenstates.
We conclude our work with a short discussion in Sec.~\ref{sec:summary}.
Extra details such as eigenoperators in generalized AKLT model and additional examples of QMBS in the spin-1 XY model are relegated to the Appendices.

\section{CORRELATION MATRIX} \label{sec:CorrMat}
Our work is primarily motivated by the information perspective on the ETH. According to this hypothesis, the diagonal matrix element of a local observable, $\hat{O}$, with respect to a thermal eigenstate $\ket{n}$ approaches the thermal ensemble average as the system size increases  \cite{Srednicki_1994ca,Srednicki_1999ta,DAlessio_2016fq}
\begin{equation}
  \mel{n}{\hat{O}}{n} \rightarrow \frac{1}{Z} \mathrm{Tr} \, e^{-\beta \hat{H}} \hat{O} \, ,
\end{equation}
where $Z = \mathrm{Tr} \, e^{-\beta \hat{H}}$ is the partition function and $\beta$ represents the inverse temperature (with $k_{B}=1$). This implies that a thermal eigenstate encodes sufficient information to infer the underlying Hamiltonian \cite{Garrison_2018da}, consistent with the fact that a wavefunction has exponentially many components, while a local Hamiltonian has only extensively many parameters. In other words, one might not be able to reconstruct the Hamiltonian from a low-entangled QMBS. These ideas are solidified by using the correlation matrix, originally introduced in \cite{Qi_2019da,Chertkov_2018ci}, as an operational tool for reconstructing the underlying local Hamiltonian from a single eigenstate.  Let $\ket{\psi}$ be a wave function, and let $\mathcal{V}$ denote a real vector space of Hermitian operators spanned by a set of operator basis $\{\hat{L}_{i} \}$ that acts on $\ket{\psi}$. For an arbitrary Hermitian operator $\hat{O} = \sum_{i} w_{i} \hat{L}_{i} \in \mathcal{V}$, we have the following inequality
\begin{equation} \label{eq:OOinequality}
	\langle \psi | \hat{O}^{2} | \psi \rangle - \big(\langle \psi | \hat{O} | \psi \rangle \big)^{2} = \langle \psi_{\perp} | \psi_{\perp} \rangle \ge 0 \, ,
\end{equation}
where $\hat{O} \ket{\psi} = \xi \ket{\psi} + \ket{\psi_{\perp}}$, and $\ket{\psi_{\perp}}$ is orthogonal to $\ket{\psi}$, i.e., $\langle \psi | \psi_{\perp} \rangle =0$. This inequality can be cast into a quadratic form
\begin{equation} \label{eq:inequality_M}
    \mathbf{w}^{T} \cdot \mathbf{M} \cdot \mathbf{w} \geq 0 \, ,
\end{equation}
where $\mathbf{w} \equiv (w_{1}, w_{2}, \cdots)^{T}$ is a vector of coefficients, and $\mathbf{M}$ is the real symmetric \emph{correlation matrix} defined as
\begin{equation} \label{eq:Mij}
  M_{ij} = \dfrac{1}{2} \mel{\psi}{(\hat{L}_{i}\hat{L}_{j} + \hat{L}_{j}\hat{L}_{i})}{\psi} - \mel{\psi}{ \hat{L}_{i}}{\psi} \! \!\mel{\psi}{ \hat{L}_{j}}{\psi} \, .
\end{equation}
The correlation matrix $\mathbf{M}$ is positive semidefinite, and equality in the quadratic form is attained only when $\mathbf{w}$ belongs to its null space, $\mathbf{w} \in \mathrm{Ker}(\mathbf{M})$. In this case, the corresponding $\hat{O}$ is referred to as an \textit{eigenoperator} of $\ket{\psi}$, meaning $\hat{O} \ket{\psi} = \xi \ket{\psi}$.
Let $N_{0} \equiv \mathrm{dim}(\mathrm{Ker}(\mathbf{M}))$ denote the dimensionality of the null space of $\mathbf{M}$.
We can then construct $N_{0}$ linearly independent eigenoperators of $\ket{\psi}$.
If the original Hamiltonian belongs to the chosen operator space $\mathcal{V}$, then the uniqueness of Hamiltonian reconstruction from thermal eigenstates is equivalent to $N_{0}=1$ \cite{Qi_2019da,Chertkov_2018ci,Garrison_2018da}, assuming the system possesses no symmetries. We shall see below that exact QMBS are characterized by an often much larger $N_{0}$, which agrees with the necessary condition of exact QMBS reached in the commutant algebra framework \cite{Moudgalya_2024ec}. Moreover, this observation is also consistent with the fact that a structured low-entangled state can often be annihilated by many local projection operators \cite{Yao_2022be}.

\section{EXACT QMBS} \label{sec:exact_QMBS}
A paradigmatic model that hosts exact QMBS is the one-dimensional Affleck--Kennedy--Lieb--Tasaki (AKLT) Hamiltonian
\begin{equation}
    \hat{H}_{0} = \sum_{n} \left[ \frac{1}{2} \hat{\mathbf{S}}_n \cdot \hat{\mathbf{S}}_{n+1} + \frac{1}{6}(\hat{\mathbf{S}}_n \cdot  \hat{\mathbf{S}}_{n+1})^2 + \frac{\hat{\mathbb{I}}}{3} \right] \, ,
\end{equation}
where $\hat{\mathbf{S}}_{n}$ is the spin-1 operator at site $n$, and $\hat{\mathbb{I}}$ is the identity operator \cite{Affleck_1987rr,Moudgalya_2018ee,Moudgalya_2018eo}. The known series of QMBS $\ket{\mathcal{S}_{2l}}$ have eigenenergies of $E_{l} = 2l$ and are captured by the SGA formalism. Their explicit form is given by
\begin{equation}
    \ket{\mathcal{S}_{2l}} = (\hat{Q}^{+})^{l} \ket{G} \, , \quad l = 0, 1, \cdots \, ,
\end{equation}
where $\ket{G}$ is the ground state of the AKLT model and $\hat{Q}^{+} = \sum_{n=1}^{L} (-1)^{n} (\hat{S}_{n}^{+})^{2}$ with $\hat{S}_{n}^{+}$ being the spin-1 raising operator at site $n$ \cite{Moudgalya_2020ep,Mark_2020us,Rozon_2024bu}.
Interestingly, this scar tower $\{ \ket{\mathcal{S}_{2l}} \}$ is stable against certain disorder perturbations: they persist as eigenstates of the following disordered Hamiltonian $\hat{H} = \hat{H}_{0} + \hat{H}_{\text{dis}}$,
\begin{equation} \label{eq:generalizedAKLT}
    \hat{H} = \hat{H}_{0} + \sum_{n,m,m'} \left[ \xi_{m,m'}^{n} (\ketbra{J_{2,m}}{J_{2,m'}})_{n,n+1} + \text{h.c.} \right] \, ,
\end{equation}
where $n$ indexes lattice sites, $m$ and $m'$ are restricted to taking $-2, -1, 0$ in the summation, $\xi_{m,m'}^{n}$ are arbitrary complex numbers, and $\ket{J_{j,m}}$ inside the parenthesis is the total spin $j$, total magnetization $m$ state of two spins at sites $n$ and $n+1$ \cite{Mark_2020us}. Using $\ket{J_{j,m}}$, the pure AKLT Hamiltonian $\hat{H}_{0}$ can be rewritten as
\begin{equation}
	\hat{H}_{0} = \sum_{n} \sum_{m=-2}^{2} (\ketbra{J_{2,m}}{J_{2,m}})_{n,n+1} \equiv \sum_{n} \hat{P}^{(2)}_{n,n+1} \, ,
\end{equation}
where $\hat{P}^{(2)}_{n,n+1}$ is the projection operator of two spins at sites $n$ and $n+1$ onto the space of total angular momentum $j=2$ \cite{Affleck_1987rr}.

\begin{figure}[htbp]
    \centering
    \includegraphics[width=0.9\linewidth]{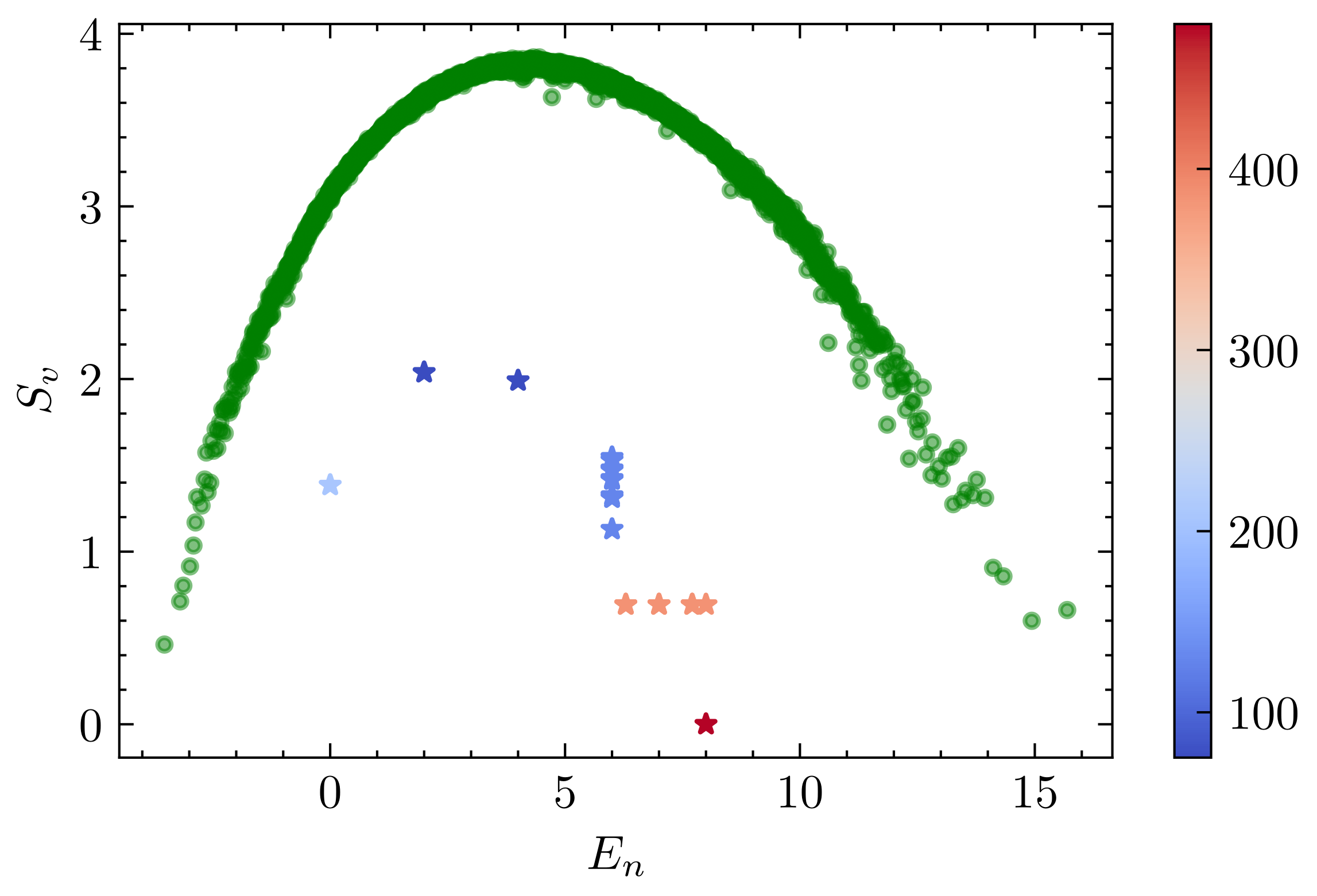}
    \caption{Plot showing the bipartite entanglement entropies of the eigenstates of the generalized AKLT Hamiltonian \eqref{eq:generalizedAKLT} against their eigenenergies $E_{n}$ for system size $L=8$. Thermal eigenstates are represented by green circles, all uniformly having $N_0 = 1$. Scar states are highlighted with stars, and their corresponding $N_0$ values are color-coded according to the color bar.}
    \label{fig:AKLT}
\end{figure}

To show that the number of correlation-matrix zeros $N_{0}$ can distinguish scar states from thermal eigenstates, we start with a simple range-2 local basis $\hat{L}_{i}$. Just as with Pauli matrices for spin-$1/2$ systems, we can use the standard generators of the SU(3) group, the eight Gell-Mann matrices $\lambda_{a}~(a=1, 2, \cdots, 8)$, together with the identity matrix $\lambda_{0}$ to define the operator basis $\hat{L}_{i}$,
\begin{equation} \label{eq:Li_AKLT}
    \hat{L}_{i} = \left( \lambda_{a} \otimes \lambda_{b} \right)_{n,n+1} \, ,
\end{equation}
where the subscript $i$ labels the operator basis and $n=1, 2, \cdots, L$ labels the lattice sites ($L+1$ is identified as $1$) so that $\hat{L}_{i}$ acts nontrivially only on sites $n$ and $n+1$.
Here $a = 0, 1, \cdots, 8$ and $b=1, 2, \cdots, 8$ so that we include all single-site and two-site operators but exclude the trivial identity operator, which makes the correlation matrix $M$ of dimension $72L \times 72 L$.
With this set of Hermitian operator basis $\{\hat{L}_{i}\}$, the correlation matrix $\mathbf{M}_{k}$ for each eigenstate $\ket{\psi_{k}}$ is then constructed according to the definition \eqref{eq:Mij}. One can then diagonalize $\mathbf{M}_{k}$ and identify the kernel space associated with zero eigenvalues (up to machine precision). Furthermore, as explained in Sec.~\ref{sec:CorrMat}, a set of linearly independent basis vectors $\{\mathbf{e}^{(j)}\}$ of the kernel space, i.e. $\operatorname{span}\{\mathbf{e}^{(1)}, \mathbf{e}^{(2)}, \cdots, \mathbf{e}^{(N_{0})}\} = \operatorname{Ker}(\mathbf{M}_{k})$ where $N_{0} \equiv \operatorname{dim}(\operatorname{Ker}(\mathbf{M}_{k}))$ can be used to construct $N_{0}$ linearly independent eigenoperators $\hat{O}^{(j)} = \sum_{i} e^{(j)}_{i} \hat{L}_{i}$ for $j=1, 2, \cdots, N_{0}$.
On the other hand, the Hamiltonian \eqref{eq:generalizedAKLT} consists of only two-body interactions and our $\hat{L}_{i}$ includes all single-site and two-site operators.
This means that $\hat{H} \in \mathcal{V} = \operatorname{span}(\hat{L}_{i})$ and $\hat{H}$ can be expanded using these operator bases $\hat{H} = \sum_{i} e_{i} \hat{L}_{i}$ for some real vector $\mathbf{e}$.
As $\hat{H}$ is an eigenoperator of each eigenstate $\ket{\psi_{k}}$ and eigenoperators are in one-to-one correspondence with the null vectors of $\mathbf{M}_{k}$, $\mathbf{e}$ must be a null vector of each $\mathbf{M}_{k}$. In other words, $N_{0} \geq 1$ for every eigenstate.
As illustrated in Fig.~\ref{fig:AKLT}, eigenstates with energies $E_{l}=0, 2, \cdots, L$ indeed exist and are numerically checked to correspond to SGA scar states $\ket{\mathcal{S}_{2l}}$.
Additionally, other scar states with anomalously low bipartite entanglement entropies are observed.
Regarding the correlation matrix spectrum, all thermal eigenstates unanimously have $N_{0}=1$, meaning the original Hamiltonian can be reconstructed from a single thermal eigenstate \cite{Qi_2019da,Chertkov_2018ci,Garrison_2018da}. In contrast, all the scar states exhibit larger values of  $N_{0}$, making the correlation-matrix zeros a reliable \textit{qualitative} probe for scar states. This stands in contrast to the \textit{quantitative} probe of entanglement entropy, which is prone to finite-size effects.

Further analysis on system sizes $L=6, 8, \cdots, 16$ shows that the minimal number of $N_{0}$ of SGA scar states satisfies $N_{0}^{(\text{min})} = 9L + 3$. This number has to be compared with the known $9L+2$ Hermitian eigenoperators of the SGA scar states: the Hamiltonian $\hat{H}_{0}$; the total magnetization $\hat{S}^{z} = \sum_{n} \hat{S}_{n}^{z}$; $9L$ Hermitian disorder terms in $\hat{H}_{\text{dis}}$ \cite{Mark_2020us}. This means there is an additional linearly independent range-2 operator, denoted as $\hat{A}$, that is an eigenoperator of all SGA scar states. Thus, the most general range-2 Hamiltonian hosting the SGA scar states is $\hat{H} + \xi \hat{A}$ where $\xi$ is an arbitrary real number. Using the Gram--Schmidt orthogonalization procedure, the explicit expression of the remaining 2-local eigenoperator $\hat{A}$ is given by
\begin{equation}
	\hat{A} = \sum_{n} (-1)^{n} \hat{A}_{n} \, ,
\end{equation}
where $n$ labels the lattice site and $\hat{A}_{n}$ in matrix representation reads
\begin{align}
	\hat{A}_{n} = & \sum_{a=1}^{8} c_{a} \left( \lambda_{a} \otimes \lambda_{a} \right)_{n,n+1}  + 2 \left( \lambda_{1} \otimes \lambda_{6} + 1 \leftrightarrow 6 \right)_{n,n+1}  \nonumber \\
	& + 2 \left( \lambda_{2} \otimes \lambda_{7} + 2 \leftrightarrow 7 \right)_{n,n+1} \nonumber \\
	& + \frac{4}{\sqrt{3}} \left( \lambda_{3} \otimes \lambda_{8} + 3 \leftrightarrow 8 \right)_{n,n+1} \, ,
\end{align}
with $c_{a} = (-11/3, -11/3, -5, 1, 1, 53/3, 53/3, 7)$ accurate to machine precision. This eigenoperator $\hat{A}$ is closely related to the recently identified staggered projection eigenoperator $\sum_{n} (-1)^{n} \hat{P}^{(2)}_{n,n+1}$ of the SGA scar states \cite{Rozon_2024bu}
\begin{align}
	\hat{A} = & \sum_{n} (-1)^{n} \Big[ 106 \hat{P}^{(2)} - 170 \ket{J_{2,-2}}\bra{J_{2,-2}} \nonumber \\ \label{eq:A_P_linear_com}
	 &-128 \ket{J_{2,-1}}\bra{J_{2,-1}} - 88 \ket{J_{2,0}}\bra{J_{2,0}} \Big]_{n,n+1} \\
	 \equiv & \sum_{n} (-1)^{n} \hat{B}_{n}
\end{align}
where a two-site operator $\hat{B}_{n}$ has been introduced to denote the four terms in the square bracket of Eq.~\eqref{eq:A_P_linear_com}, and the detailed derivation is presented in Appendix A.  The above equality $\sum_{n} (-1)^{n} \hat{A}_{n} = \sum_{n} (-1)^{n} \hat{B}_{n}$ can be understood by noting that the local difference
\begin{align}
	\hat{A}_{n} -\hat{B}_{n} = & 3\left[ (\hat{S}_{n}^{z})^{2} + (\hat{S}_{n+1}^{z})^{2} \right] \nonumber \\
     & -39(\hat{S}_{n}^{z}+ \hat{S}_{n+1}^{z}) - 20 \hat{I}_{n,n+1} \, ,
\end{align}
where $\hat{I}_{n,n+1}$ is the identity operator on sites $n$ and $n+1$, vanishes upon a $(-1)^{n}$ summation. \\


\section{SCAR EMBEDDING AND COMPATIBILITY} \label{sec:embedding}
\begin{figure}[tbph]
    \centering
    \includegraphics[width=0.9\linewidth]{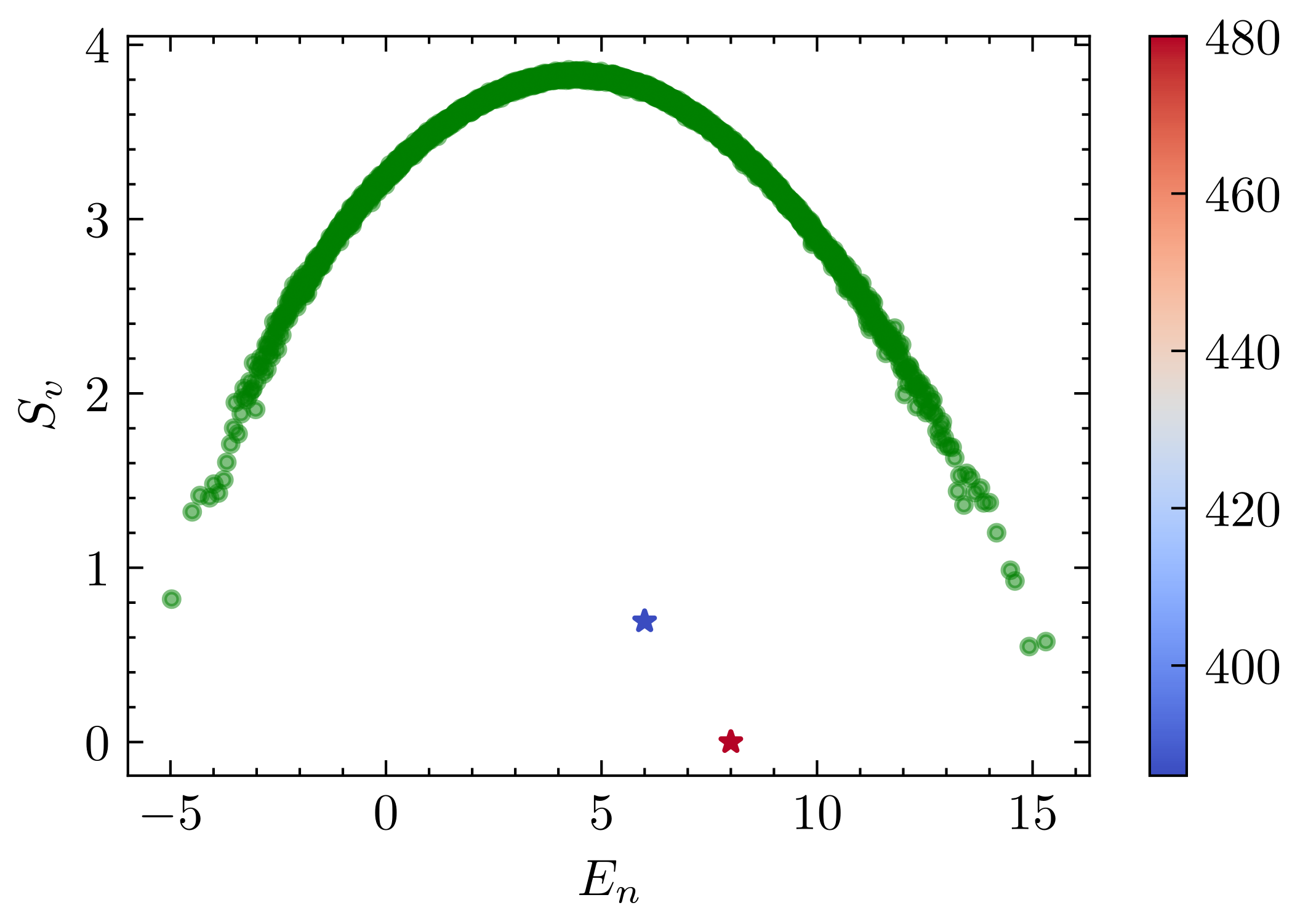}
    \caption{Plot of bipartite entanglement entropies against eigenenergies $E_{n}$ of a Hamiltonian $\hat{H}'$ that embeds the $\ket{1_{k=\pi}}$ and $\ket{F}$ states as the only scar states for system size $L=8$. The scar states are represented by stars, with their colors corresponding to the values of $N_0$ as indicated by the color bar.}
    \label{fig:AKLT_embed}
\end{figure}

In addition to providing a qualitative diagnosis of scar states, the correlation matrix framework enables flexible manipulation of the scar state subspace and offers insight into its internal structure. Let $\mathcal{A}_{\ket{\psi}}$ denote the \textit{eigenoperator space} of $\ket{\psi}$, which is the operator space spanned by $\{ \hat{A} = \sum_{i} w_{i} \hat{L}_{i}\}$, where $\mathbf{w}$ is a null eigenvector of the correlation matrix $\mathbf{M}$ constructed from $\ket{\psi}$. Given two eigenstates $\ket{\psi}$ and $\ket{\phi}$ of a Hamiltonian $\hat{H}$, if $\mathcal{A}_{\ket{\psi}} \subset \mathcal{A}_{\ket{\phi}}$, then we can always select an operator $\hat{A} \in \mathcal{A}_{\ket{\phi}}$ but $\hat{A} \notin \mathcal{A}_{\ket{\psi}}$ and add it to the original Hamiltonian to construct a new Hamiltonian,
\begin{equation}
    \hat{H}' = \hat{H} + \hat{A} \, ,
\end{equation}
of which $\ket{\phi}$ but not $\ket{\psi}$ remains an eigenstate. To illustrate this, we use the generalized AKLT model \eqref{eq:generalizedAKLT} as an example. Scar states in this model include the ferromagnetic state $\ket{F} = \ket{11\cdots 11}$, single-magnon states \cite{Moudgalya_2018ee,Tang_2022mq},
\begin{equation}
    \ket{1_{k}} = \sum_{n} e^{ikn} \big\vert \underbrace{1\cdots 1 }_{n-1} 0 \underbrace{1 \cdots 1}_{L-n} \big\rangle \, ,
\end{equation}
where $n=0,1,\cdots, L-1$ in the summation and $k=2\pi m/L$ ($m=0,1,\cdots, L-1$) is the quasimomentum, and several other families such as the SGA tower.
A detailed analysis of the eigenoperator spaces reveals that,
\begin{equation}
    \mathcal{A}_{\ket{1_{k}}} \subset \mathcal{A}_{\ket{F}} \, , \quad \forall \, k
\end{equation}
and that $\mathcal{A}_{\ket{1_{k}}}$ is not a subset of the eigenoperator space of any other scar state.
Moreover, although the dimension of $\mathcal{A}_{\ket{1_{k}}}$ is the same for all $k$, their individual eigenoperator spaces are different.
This means that we can construct a Hamiltonian to retain any given $\ket{1_{k}}$ and $\ket{F}$ as the only scar states. We can also add terms to make $\ket{F}$ the unique scar state.
However, for a Hamiltonian with on-site and nearest-neighbor interactions, if $\ket{1_{k}}$ is a scar state, $\ket{F}$ is necessarily an eigenstate.
As shown in Fig.~\ref{fig:AKLT_embed}, we can construct a Hamiltonian $\hat{H}'= \hat{H} + \sum_{i} \xi_{i} \hat{A}_{i}$ where $\hat{A}_{i} \in \mathcal{A}_{\ket{1_{\pi}}}$ and $\xi_{i}$ are random numbers to embed the scar space $\mathcal{S} = \operatorname{span}\{ \ket{1_{\pi}}, \ket{F} \}$.  Similarly, since the eigenoperator space $\mathcal{A}_{\ket{G}}$ is not a subset of any other scar state, we can construct a Hamiltonian with $\ket{G}$ as the only scar state \footnote{Refer to Appendix B for an in-depth analysis of the scar eigenoperator spaces and scar embedding examples.}.

\section{UNIFICATION AND GENERALITY} \label{sec:unification}
Our method of using correlation-matrix zeros $N_{0}$ to probe exact QMBS unifies many mathematical frameworks. To illustrate the generality of our approach, we consider three prominent examples: the Shiraishi--Mori framework, the non-Abelian Lie algebra construction, and the SGA formalism.
In the Shiraishi--Mori formalism \cite{Shiraishi_2017sc}, the scarring Hamiltonian takes the following form
\begin{equation}
    \hat{H} = \sum_{n} \hat{P}_{n} \hat{h}_{n} \hat{P}_{n} + \hat{H}' \, ,
\end{equation}
where $\hat{P}_{n}$ are a set of local projectors that annihilate the scar subspace $\mathcal{S}$, $\hat{h}_{n}$ are some local Hermitian operators, and $\hat{H}'$ is some Hamiltonian that satisfies $[\hat{P}_{n}, \hat{H}'] \mathcal{S} =0$. It follows that for any scar state $\ket{\psi} \in \mathcal{S}$, the projector $\hat{P}_{n}$ is an eigenoperator of $\ket{\psi}$ and each $\hat{P}_{n}$ contributes a zero eigenvector to the correlation matrix $\mathbf{M}$ for an appropriately large operator space $\mathcal{V}$.
This means that for any scar state $N_{0} \geq N_{p}$ where $N_{p}$ is the number of projectors $\hat{P}_{n}$, which distinguishes it from thermal eigenstates. Similarly, in the ``tunnels to towers'' construction \cite{ODea_2020ft}, the scarring Hamiltonian takes the following form
\begin{equation}
    \hat{H} = \hat{H}_{\text{sym}} + \hat{H}_{\text{SG}} + \hat{H}_{\text{A}} \, ,
\end{equation}
where $\hat{H}_{\text{sys}}$ is a Hamiltonian with non-Abelian symmetry to endow the spectrum with a degenerate subspace, $\hat{H}_{\text{SG}}$ consists of a linear combination of generators in the Cartan subalgebra to lift the degeneracy, and a final Hamiltonian $\hat{H}_{\text{A}}$ that annihilates a targeted subspace $\mathcal{S}$ is needed to break symmetries and promote $\mathcal{S}$ to a scar subspace. In this case, $\hat{H}_{\text{A}}$ is also an eigenoperator for the scar states, making it detectable via the correlation matrix approach.

To demonstrate that our method also captures QMBS in the SGA formalism \cite{Moudgalya_2020ep}, it is convenient to use the equivalent Mark--Lin--Motrunich (MLM) framework \cite{Mark_2020us}
\begin{equation} \label{eq:rSGA}
    \left( [\hat{H}, \hat{Q}^{+}] - \epsilon \hat{Q}^{+} \right) \mathcal{S} = 0 \, ,
\end{equation}
where $\mathcal{S}$ is the scar subspace, $\hat{H}$ is the Hamiltonian, and $\hat{Q}^{+}$ is the ladder operator that repeatedly acts on a reference eigenstate $\ket{\psi_{0}}$ to generate the entire scar tower $\{ \ket{\mathcal{S}_{2l}} \equiv (\hat{Q}^{+} )^{l} \ket{\psi_{0}} \}$. Therefore, as long as $\hat{H}$ and $\hat{Q}^{+}$ are sums of local operators with maximum range $k$, the left-hand side, denoted as $\hat{A}$, of \eqref{eq:rSGA} is a sum of local operators. Although $\hat{A}$ is in general not Hermitian, this can be easily remedied by choosing the operator basis $\{\hat{L}_{i}\}$ to be non-Hermitian and subsequently modifying the form of the correlation matrix $\mathbf{M}$ as
\begin{equation}
    M_{ij} =
    \bra{\psi} \hat{L}_{i}^{\dagger} \hat{L}_{j} \ket{\psi} - \bra{\psi} \hat{L}_{i}^{\dagger} \ket{\psi} \bra{\psi} \hat{L}_{j} \ket{\psi} \, .
\end{equation}
The correlation matrix $\mathbf{M}$ is then a positive semidefinite Hermitian matrix, and the annihilator $\hat{A}$ is also in one-to-one correspondence with a null eigenvector of $\mathbf{M}$. In fact, as demonstrated in the generalized AKLT model, even using a Hermitian operator basis with a real symmetric $\mathbf{M}$ is often sufficient in detecting scar states.

Our method is also closely related to the commutant algebra method proposed in a recent work \cite{Moudgalya_2024ec}. Given a set of QMBS $\{\ket{\psi_{s}}\}$ and the corresponding $\dagger$-algebra
\begin{equation}
    \mathcal{C} = \langle\!\langle \{ \ket{\psi_{s}}\!\bra{\psi_{s'}}\} \rangle\!\rangle
\end{equation}
where $\langle\!\langle \cdots \rangle\!\rangle$ denotes the associative algebra generated through products, Hermitian conjugations, and linear combinations with complex coefficients, the first task is to identify a series of local Hermitian operators (strictly local) $\{\hat{H}_{\alpha}\}$ or sums of local Hermitian operators (extensively local) such that each of them is an eigenoperator of every $\ket{\psi_{s}}$. The main point of the commutant algebra formalism is that in many cases a set of $\{\hat{H}_{\alpha}\}$ can be found such that $\mathcal{C}$ is the centralizer (commutant) of the $\dagger$-algebra $\mathcal{A} \equiv \langle\!\langle \{\hat{H}_{\alpha}\} \rangle\!\rangle$. A natural method to search for these operators $\hat{H}_{\alpha}$ is the correlation matrix method, where strictly local $\hat{H}_{\alpha}$ can be identified by restricting operator bases $\hat{L}_{i}$ to act only nontrivially in a local region \cite{Yao_2022be}.
Since the commutant algebra treats conventional symmetries and QMBS on a similar footing without taking thermal eigenstates into consideration, the existence of strictly local or extensive local eigenoperators is only a necessary condition in the commutant algebra framework.
On the contrary, our approach involves a comparison of $N_{0}$ between QMBS and thermal eigenstates, which is unaffected by the presence of conventional symmetries. Thus, it serves as a sufficient condition or even a definition of exact QMBS.
In our study, all thermal eigenstates uniformly have the same value of the correlation-matrix zero $N_{0}$, from which we can uniquely construct the underlying Hamiltonian $\hat{H}$, modulo the addition of extra conserved quantities such as total magnetization and a constant scaling factor.
Our result of $N_{0}$ for thermal eigenstates is thus consistent with the implication of the ETH that a thermal eigenstate encodes enough information to determine the underlying Hamiltonian \cite{Garrison_2018da,Qi_2019da,Chertkov_2018ci}.
On the contrary, a larger value of $N_{0}$ for an exact QMBS implies that, even after accounting for extra conserved quantities and a constant rescaling, it can still be an eigenstate of many distinct Hamiltonians $\hat{H}_{1}, \hat{H}_{2}, \cdots$. The eigenstate expectation value of a local observable $\hat{O}$ is then naturally different from $\operatorname{Tr}' (e^{-\beta \hat{H}} \hat{O}) /\operatorname{Tr}' e^{-\beta \hat{H}}$ where $\operatorname{Tr}'$ designates the evaluation in a given symmetry sector, resulting in a violation of the ETH \cite{Moudgalya_2024ec}. It is from this information perspective of the ETH that we propose using a larger value of $N_{0}$ as a definition of exact QMBS.
Another notable difference is that we study the eigenoperator space of each $\ket{\psi_{s}}$ separately instead of identifying their common intersection $\{\hat{H}_{\alpha}\}$, which gives us an advantage in manipulating the scar subspaces and deriving compatibility conditions for embedding different QMBS.

\section{APPROXIMATE QMBS} \label{sec:approximate_QMBS}
\begin{figure*}[tbph]
    \centering
    \includegraphics[width=0.975\linewidth]{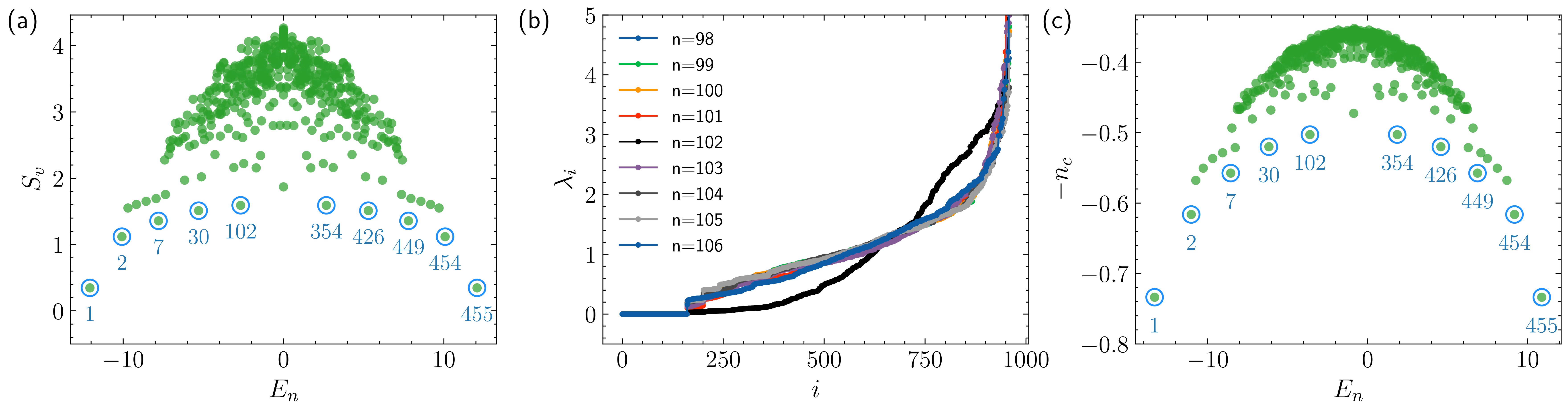}
    \caption{
        (a) Bipartite von Neumann entanglement entropy $S_{v}$ of eigenstates $E_{n}$ plotted against the eigenenergies of the PXP model for $L=20$ in the $(k, I) = (0, 1)$ sector, where $k$ is the lattice momentum and $I$ is the parity quantum number. The QMBS are marked with blue circles with annotated numbers below designating the eigenstate index $n$ as eigenenergy increases.
        (b) Correlation matrix spectra of operator range $r=3$ for the scar state $n =102$ (black dots) and the surrounding thermal eigenstates (colored dots).
        (c) Plot of $-n_{c}$ against the eigenenergies of the PXP model for $L=20$ and $(k, I)=(0, 1)$. The blue circles and annotated text have the same meaning as in panel (a).
    }
    \label{fig:PXP_Sv_Nc}
\end{figure*}

For the several paradigmatic exact QMBS we have tested \footnote{Refer to Appendix C for extra examples of QMBS in the spin-1 XY model which also includes references \cite{Kitazawa_2003as}.}, all scar states distinguish themselves from thermal eigenstates by having a much larger $N_{0}$. However, for approximate scars that have neither known analytic expressions nor are captured by previous mathematical formalisms, the single number $N_{0}$ may be inadequate. A prime example of this is the well-known PXP model, described by the following Hamiltonian \cite{Turner_2018we,Turner_2018qs}
\begin{equation}
    \hat{H}=\sum_{n} \hat{P}_{n-1} \hat{X}_{n} \hat{P}_{n+1} \, ,
\end{equation}
where $\hat{P}_{n}= (1 - \hat{Z}_{n})/2$ is the projection operator at site $n$, and  $\hat{X}_{n}$ and $\hat{Z}_{n}$ are standard Pauli operators. Except for special eigenstates in the middle of the spectrum, the analytic expressions of QMBS are, in general, unavailable \cite{Lin_2019eq,Ivanov_2025ve}.
In our correlation matrix study, we choose the matrix representation of range-$r$ operator bases $\hat{L}_{i}$ acting on lattice sites $n+1, n+2, \cdots, n+r$ as
\begin{equation}
    \hat{L}_{i} = (\sigma_{a_{1}} \otimes \sigma_{a_{2}} \cdots \otimes \sigma_{a_{r}})_{n} \, ,
\end{equation}
where $\sigma_{a_{j}}$ denotes the identity and the three Pauli matrices $(X, Y, Z)$ for $a_{j}=0$ and $a_{j}=1, 2, 3$, respectively. As before, $a_{1}$ is restricted to $1,2,3$ while other $a_{i}$ take $0,1,2,3$. This ensures that our operator bases exclude the trivial identity operator but includes all nontrivial local operators up to range $r$. We employ periodic boundary conditions and use a system size of $L=20$ in our study.
The set of approximate QMBS can be easily identified by anomalously low bipartite entanglement entropies of the eigenstates, which are circled in blue in Fig.~\ref{fig:PXP_Sv_Nc}(a). We then pick a representative scar state with eigenenergy index $n=102$, study its correlation matrix spectrum (eigenvalue spectrum of the correlation matrix) and compare it with the surrounding thermal eigenstates.
As shown in Fig.~\ref{fig:PXP_Sv_Nc}(b), the scar state (colored black) has the same $N_{0}$ as that of the thermal eigenstates, which is associated with the Rydberg blockade condition. However, the correlation matrix spectrum reveals marked differences between the scar states and the thermal states through the presence of many approximate zero eigenvalues
\footnote{Refer to Appendix D for detailed analysis of the approximate eigenoperators corresponding to approximate zeros of the scar states.}.
To quantify these differences, we introduce the following measure analogous to the role of entanglement entropy in quantifying quantum entanglement
\begin{equation}
    n_{c} = \langle e^{-\lambda_{i}} \rangle \equiv \frac{1}{N}\sum_{i=1}^{N} e^{-\lambda_{i}} \, ,
\end{equation}
to characterize the correlation matrix spectrum, where the summation over $\lambda_{i}$ is restricted to nonzero values and $N$ is the total number of nonzero $\lambda_{i}$.
As can be seen from Fig.~\ref{fig:PXP_Sv_Nc}(c), the introduced quantity $n_{c}$ effectively distinguishes scar states from thermal eigenstates. Moreover, $n_{c}$ is less sensitive to finite-size effects compared to entanglement entropy, making it a more reliable probe.

\section{IMPLICATIONS FOR THERMAL EIGENSTATES} \label{sec:thermal}
From previous discussions, for a system lacking either local symmetry (e.g., lattice gauge models) or local constraints (e.g., the PXP model), a sufficient condition for an eigenstate $\ket{\psi}$ to be nonthermal is the presence of a local operator that annihilates the state. This observation has important ramifications for the structure of thermal eigenstates. Consider the reduced density matrix $\hat{\rho}_{A} = \operatorname{Tr}_{B} \ket{\psi} \! \bra{\psi}$ of a bipartition of the system $[A : B]$, where the size of the subsystem $A$ is bounded. The reduced density matrix can be diagonalized in a local basis supported on subsystem $A$:
\begin{equation}
    \hat{\rho}_{A} = \sum_{i=1}^{d} p_{i} \ket{i} \!\bra{i} \, ,
\end{equation}
where $d$ is the Hilbert space dimension of the subsystem $A$ and $p_{i} \geq 0$ are arranged in descending order $p_{1} \geq p_{2} \geq \cdots p_{d}$. If $\hat{\rho}_{A}$ is not of full rank, i.e., $p_{k}=0$ for $k \ge \chi$ for some $\chi$, we can then construct a series of local projectors
\begin{equation}
    \hat{P}_{k} = \ket{i_{k}} \!\bra{i_{k}} \, , \quad k \geq \chi
\end{equation}
that annihilate the state $\ket{\psi}$, which is most easily seen from the Schmidt decomposition of $\ket{\psi}$. Thus, we conclude that the reduced density matrix of a thermal eigenstate for a subsystem with finite dimension must be of full rank. As a corollary, a matrix product state (MPS) with finite bond dimension cannot be thermal \cite{Moudgalya_2020lc,Yao_2022be,Moudgalya_2024ec}.

\section{SUMMARY} \label{sec:summary}
We have introduced a new method to detect quantum many-body scars (QMBS) using the correlation matrix spectrum. For exact QMBS, we demonstrate that the integer $N_0$, the number of zero eigenvalues, can \textit{qualitatively} distinguish scar states from thermal eigenstates. In this regard, $N_0$ functions similarly to how symmetry and topological numbers are used to classify phases of matter.
Our method not only unifies several previous frameworks of QMBS but is also quite accessible and flexible.
It enables the straightforward derivation of otherwise obscure analytic results, manipulation of the scar subspace, and insights into its internal constraints.
For approximate QMBS where $N_{0}$ may fall short, one can introduce quantities such as $n_{c}$ to characterize scar states, leveraging the presence of multiple approximate zero eigenvalues in the spectrum. Given the versatility of the correlation matrix approach, we anticipate that its applications will extend beyond the study of QMBS.

\section*{ACKNOWLEDGMENTS}
We thank Hui Zhai, Lei Pan, and Shang Liu for numerous useful discussions. We also thank Shang Liu for carefully reading our manuscript and valuable feedback. We acknowledge support by National Natural Science Foundation of China Grant No.~12304288 (Z. Y.), No.~12247101 (Z. Y.), and No.~12374477 (P. Z.).

\section*{APPENDIX A: RELATIONSHIP BETWEEN $\hat{A}$ AND THE STAGGERED PROJECTION EIGENOPERATOR}
Since the operator basis $\{\hat{L}_{i}\}$ used in studying the generalized AKLT model contains all possible single- and two-body local interactions (excluding the trivial identity operator), the staggered projection eigenoperator $\sum_{n} (-1)^{n} \hat{P}^{(2)}_{n,n+1}$, being a sum of two-body interactions, must belong to the $(9L+3)$-dimensional common eigenoperator space. Thus, it can be obtained by a linear combination of our $\hat{A}$ and other eigenoperators such as $(\ket{J_{2,m}}\bra{J_{2,m'}})_{n,n+1}$. As the coefficients in Eq.~\eqref{eq:A_P_linear_com} are not easy to guess, we explain here how Eq.~\eqref{eq:A_P_linear_com} is obtained in our numerical study.

Let $\mathbf{W}$ denote the $9L+3$ dimensional common eigenoperator space of the SGA scars obtained in our numerical study, let $\mathbf{w}(\hat{O})$ denote the expansion vector $\mathbf{w}$ of a 2-local (or sum of 2-local operators) $\hat{O}$ in terms of the basis operators $\hat{L}_{i}$
\begin{equation}
	\hat{O} \cong \sum_{i} w_{i} \hat{L}_{i} \, ,
\end{equation}
where $\cong$ stands for equality modulo identity operator. We first verified that the vector space spanned by $9L+3$ expansion vectors of the basis operators $\hat{H}_{0}$, $\sum_{n} (-1)^{n} \hat{P}^{(2)}_{n,n+1}$, $\hat{S}^{z}$, and $9L$ $(\ket{J_{2,m}}\!\bra{J_{2,m'}})_{n,n+1}$ is equal to $\mathbf{W}$ through the Gram--Schmidt orthogonalization procedure. Therefore, $\mathbf{w}(\hat{A})$ must be linearly dependent with the $9L+3$ expansion vectors of the basis operators mentioned above
\begin{equation*}
	\mathbf{w}(\hat{A}) = \sum_{i} c_{i} \mathbf{w}_{i}\, ,
\end{equation*}
with $i=1, \cdots, 9L+3$ labeling the basis operators. To facilitate analysis, we can write the above equation as a matrix equation
\begin{equation}
	\mathbf{w}(\hat{A}) = V \mathbf{c} \, ,
\end{equation}
where $V =[\mathbf{w}_{1}, \cdots, \mathbf{w}_{9L+3}]$ and $\mathbf{c} = (c_{1}, \cdots, c_{9L+3})^{T}$. To solve it, we employ the QR decomposition $V = QR$ and the matrix equation becomes
\begin{equation}
	\tilde{R} \mathbf{c} = \tilde{\mathbf{w}} \, ,
\end{equation}
where $\tilde{R}$ is the $(9L+3) \times (9L+3)$ upper part of $R$ and $\tilde{w}$ is a $9L+3$ dimensional vector
\begin{equation*}
	R =\left[
	\begin{array}{c}
		\tilde{R} \\ \hline
		\rule{0cm}{15pt} \text{\large 0}
	\end{array}
	\right], \qquad
	Q^{T} \mathbf{w}(\hat{A})
	= \left[ \begin{array}{c}
		\tilde{\mathbf{w}} \\ \hline
		\rule{0cm}{15pt} \text{\large 0}
	\end{array}
	\right].
\end{equation*}
The final solution of the coefficient vector $\mathbf{c}$ is then given by $\mathbf{c} = \tilde{R}^{-1} \tilde{\mathbf{w}}$, and we also verified that Eq.~\eqref{eq:A_P_linear_com} holds for the obtained $\mathbf{c}$.

\section*{APPENDIX B: SCAR EIGENOPERATOR SPACES IN GENERALIZED AKLT MODEL}
\begin{figure}[tbph]
    \centering
    \includegraphics[width=0.85\linewidth]{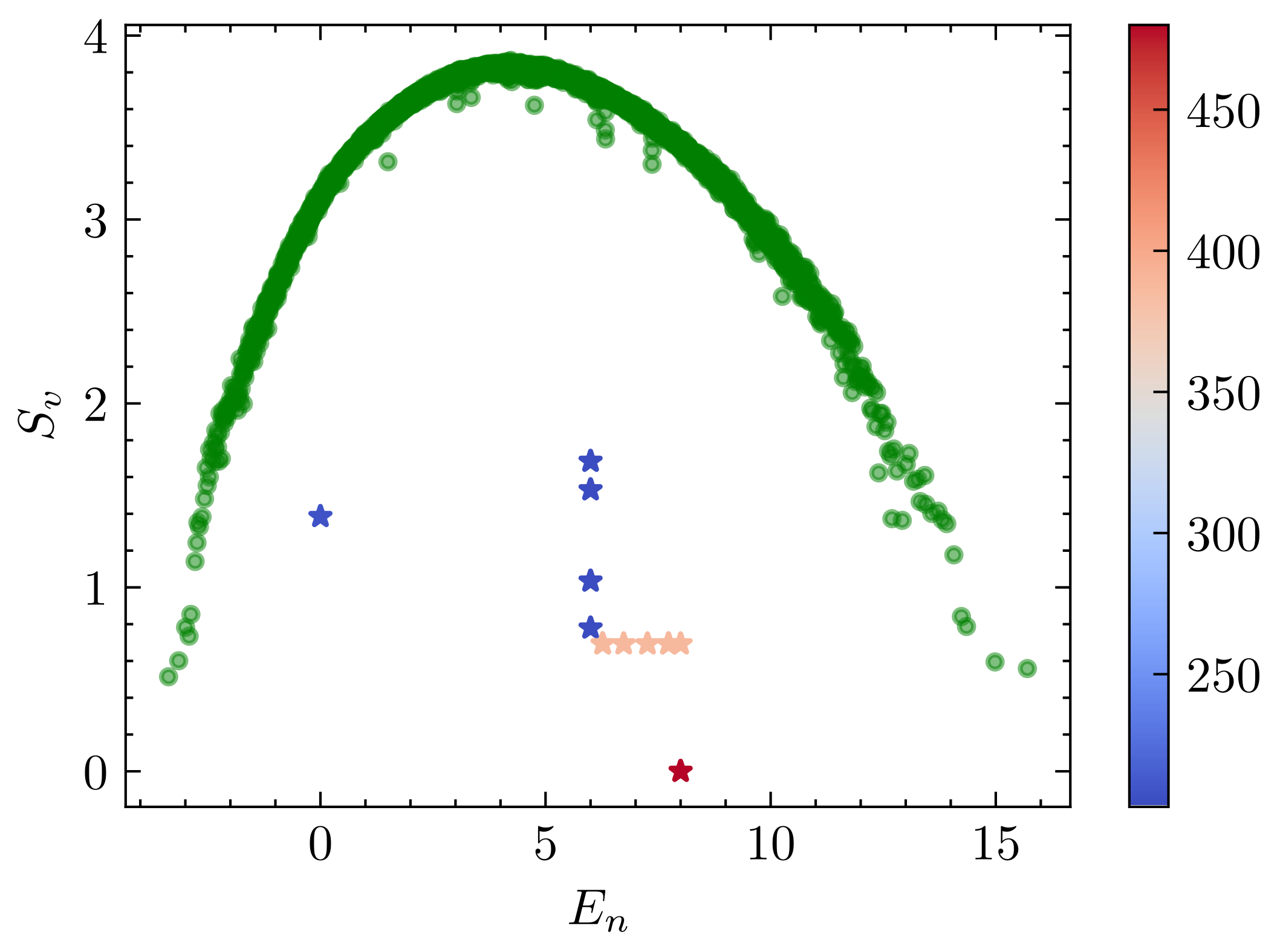}
    \caption{Plot of bipartite entanglement entropies against the eigenenergies $E_{n}$ of the Hamiltonian $\hat{H}' = \hat{H} + c \hat{A}$ for $L=8$, where $\hat{H}$ is the generalized AKLT Hamiltonian and $c$ is a real coefficient. As before eigenstates with $N_{0}=1$ are plotted as green circles, and scar states with $N_{0}>1$ are drawn as stars, color-coded by their values according to the color bar.}
    \label{fig:nzero_Sv_L8_kill_middle_SGA}
\end{figure}

In this section, we provide further details on the eigenoperator spaces of QMBS in the generalized AKLT model \eqref{eq:generalizedAKLT}. As mentioned in the main text, the minimal $N_{0}$ of the SGA scars is $9L+3$. Specifically, for tested system sizes ranging from $L = 6$ to $L = 16$, the correlation matrices of the scar states $\ket{\mathcal{S}_{2}}, \ket{\mathcal{S}_{4}}, \cdots, \ket{\mathcal{S}_{L-4}}$ have a common null space of dimension $9L+3$.
Moreover, this space is a proper subset of the intersection of the null spaces of the other SGA scars, $\ket{\mathcal{S}_{0}}, \ket{\mathcal{S}_{L-2}}, \ket{\mathcal{S}_{L}}$, which is of dimension $9L+4$. This means that there exists an eigenoperator $\hat{A}$ that can be added to the Hamiltonian to destroy the middle SGA scars $\ket{\mathcal{S}_{2}}, \ket{\mathcal{S}_{4}}, \cdots, \ket{\mathcal{S}_{L-4}}$. This operator $\hat{A}$ is a sum of local operators
\begin{equation}
    \hat{A} = \sum_{n} (-1)^{n(L-2)/2} \hat{A}_{n,n+1} \, ,
\end{equation}
and the expression of $\hat{A}_{n,n+1}$ depends on whether $L/2$ is even or odd. Omitting the subscript $(n,n+1)$ for simplicity, we have
\begin{equation}
    \hat{A}_{n, n+1} = \lambda_{1} \otimes \lambda_{2} + \lambda_{7} \otimes \lambda_{6} - \lambda_{2} \otimes \lambda_{1} - \lambda_{6} \otimes \lambda_{7} \, ,
\end{equation}
for odd $L/2$, and
\begin{align}
    \hat{A}_{n,n+1} = \; & 3\sqrt{2} \big(\ketbra{J_{0,0}}{J_{2,0}} + \text{h.c.} \big) - 7 \ketbra{J_{1,-1}}{J_{1,-1}} \nonumber \\
    & + 3 \ketbra{J_{1,0}}{J_{1,0}} + 4 \ketbra{J_{1,1}}{J_{1,1}} + 11 \ketbra{J_{2,1}}{J_{2,1}}  \nonumber \\
    & - 11 \ketbra{J_{2,2}}{J_{2,2}} \, .
\end{align}
for even $L/2$. Although we have used $\ket{m_{1}m_{2}}$ and $\ket{J_{j, m}}$ to make the expression as simple as possible, the final expressions remain cumbersome and difficult to interpret. Conversely, these intricate expressions demonstrate the strength and flexibility of our method. As shown in Fig.~\ref{fig:nzero_Sv_L8_kill_middle_SGA}, the middle SGA scars $\ket{\mathcal{S}_{2}}, \ket{\mathcal{S}_{4}}, \cdots, \ket{\mathcal{S}_{L-4}}$ are indeed destroyed when $c \hat{A}$ ($c\neq 0$) is added to the original Hamiltonian.

\begin{figure}[tbph]
    \centering
    \includegraphics[width=0.85\linewidth]{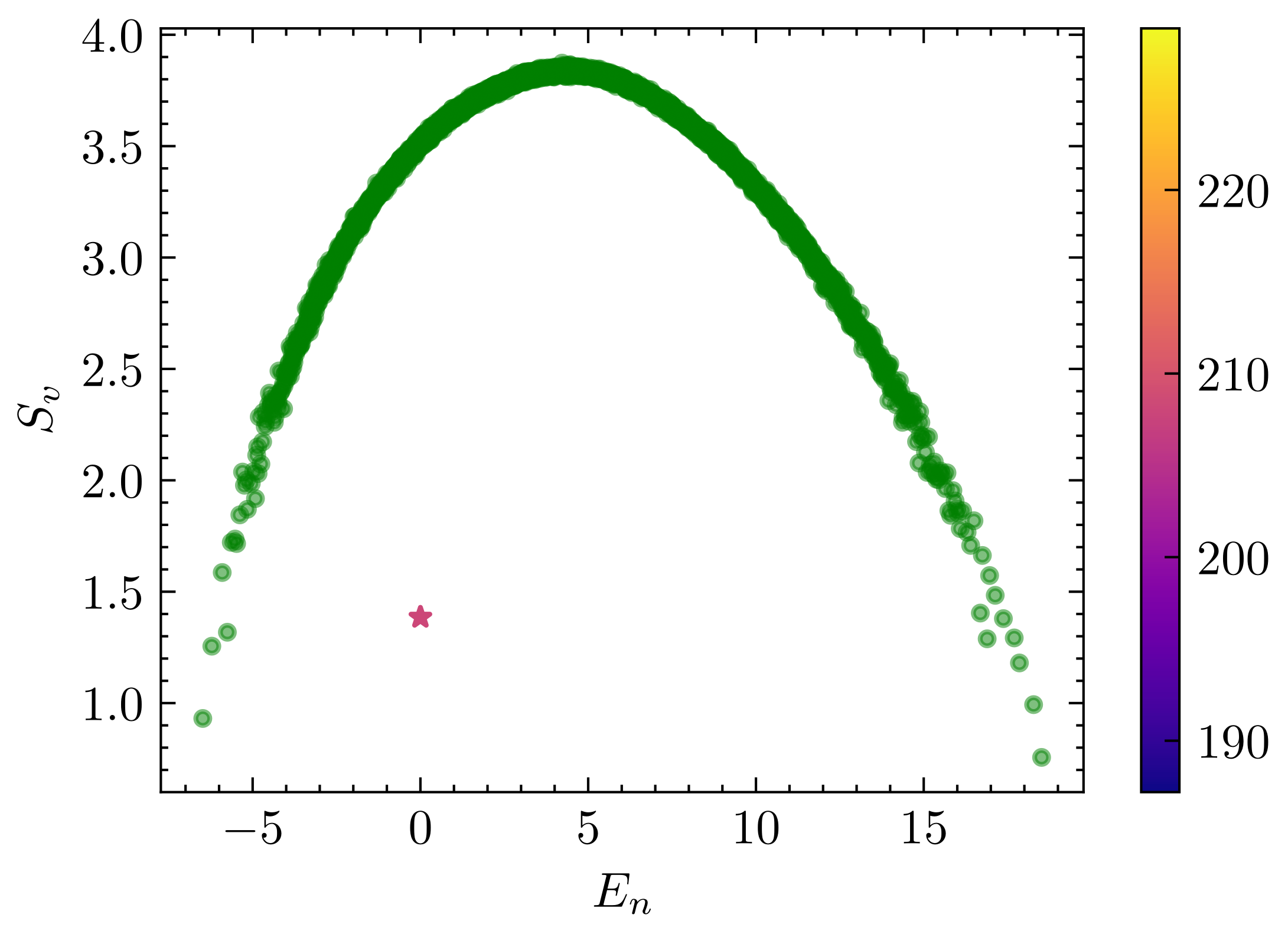}
    \caption{Plot of eigenenergies $E_{n}$ against the bipartite entanglement entropies for the embedding Hamiltonian with system size $L=8$. Thermal eigenstates with $N_{0}$ are represented as green circles, while scar states with $N_{0} > 1$ are shown as stars, color-coded by their values according to the color bar. We end up with a single scar state, the AKLT ground state $\ket{G}$, with $N_{0}=208$ for the embedding Hamiltonian \eqref{eq:H_embed_G}.}
    \label{fig:AKLT_embed_G}
\end{figure}

The ferromagnetic state $\ket{F}$, owing to its simplicity, has the largest eigenoperator space dimension. On the other hand, the ground state of the AKLT Hamiltonian, $\ket{G} \equiv \ket{\mathcal{S}_{0}}$, is unique in that its null space is not a subset of any other scar state, including the ferromagnetic state. This implies that we can construct a Hamiltonian $\hat{H}'$ of the form
\begin{equation} \label{eq:H_embed_G}
    \hat{H}' =  \hat{H} + \sum_{i} c_{i} \hat{A}_{i} \, ,
\end{equation}
where $c_{i}$ are random real numbers and $\hat{A}_{i}$ are eigenoperators of $\ket{G}$ that embed $\ket{G}$ as the only scar state. As shown in Fig.~\ref{fig:AKLT_embed_G}, such an embedding Hamiltonian successfully isolates $\ket{G}$ as the only scar state.

\section*{APPENDIX C: EXACT SCARS IN THE SPIN-1 XY MODEL}
\begin{figure}[tbph]
    \centering
    \includegraphics[width=0.85\linewidth]{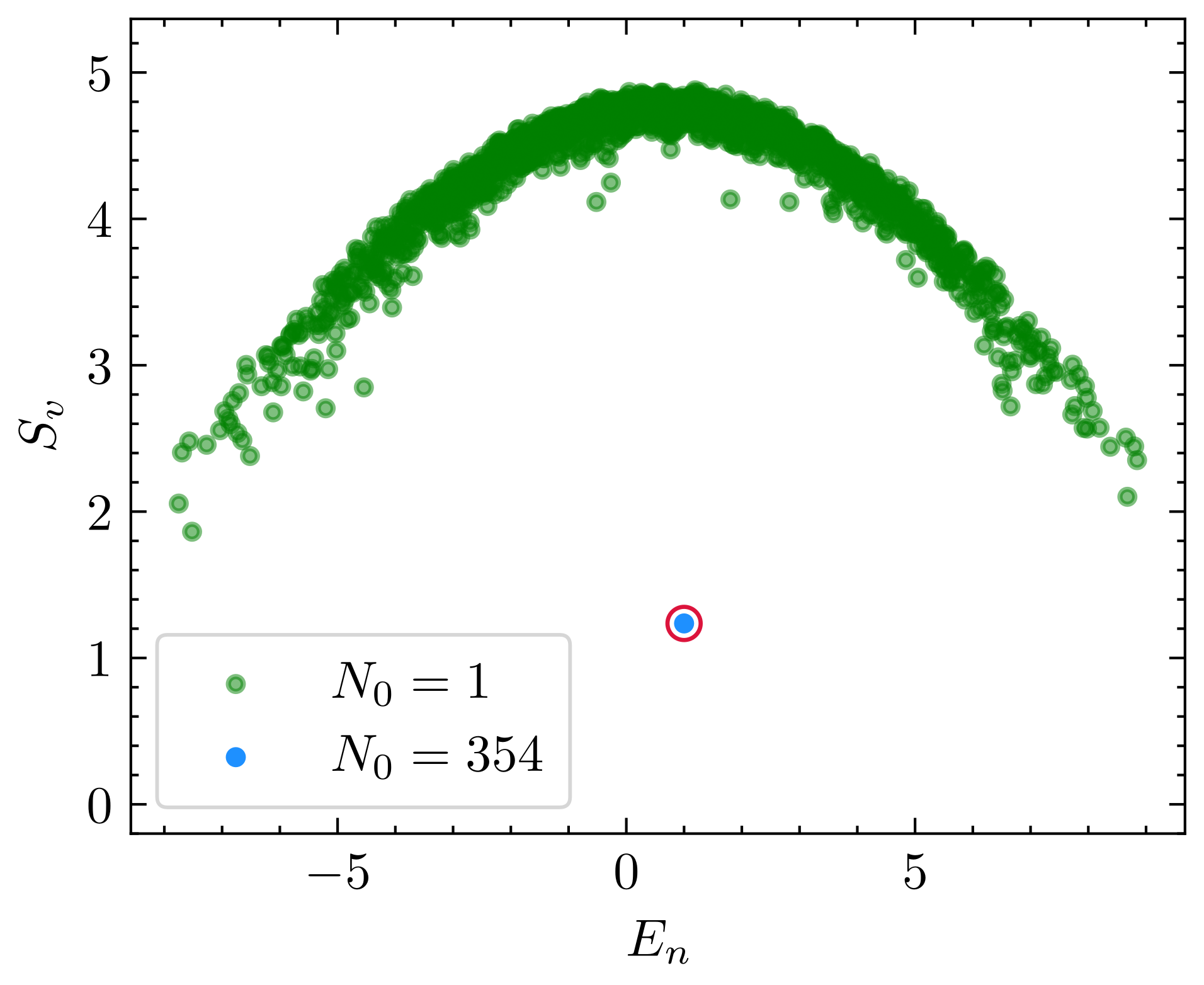}
    \caption{Plot of bipartite entanglement entropies against the eigenenergies $E_{n}$ of the spin-1 XY model in the total magnetization $M_{z}=0$ and both spatial reflection and spin inversion antisymmetric sector for system size $L=10$, $h=1.0$, $D=0.1$, and $J_{3} = 0.1$. Eigenstates with the correlation-matrix zero $N_{0}=1$ are marked as green circles. The scar state with $N_{0}=354$ is circled in red.
    }
    \label{fig:Spin1_XY_OBC}
\end{figure}

We next apply our method to the one-dimensional spin-1 XY model whose Hamiltonian is given by
\begin{align}
    \hat{H} = & \sum_{\langle ij\rangle} \left(\hat{S}_i^x \hat{S}_j^x + \hat{S}_i^y \hat{S}_j^y \right) + h \sum_i \hat{S}_i^z + D \sum_i (\hat{S}_i^z)^2 \nonumber \\
    \label{eq:Spin1_XY_OBC}
    & + J_{3} \sum_{i} (\hat{S}_i^+ \hat{S}_{i+3}^- + \text{h.c.}) \, ,
\end{align}
where $\hat{S}_{i}^{\alpha}$ ($\alpha=x, y, z$) are spin-1 operators at site $i$, $\langle ij\rangle$ denote nearest neighbors, $h$ is the external magnetic field strength, and $D$ is the anisotropy parameter. The last term, which couples spins at sites $i$ and $i+3$,  is included to break a nonlocal SU(2) symmetry under open boundary conditions (OBC) \cite{Kitazawa_2003as}. We shall first focus on OBC, where the scar states $\ket{\mathcal{S}_{l}}$ admit the following analytic expression,
\begin{equation}
    \ket{\mathcal{S}_{l}} = \mathcal{N}_{l} (\hat{J}^{+})^{l} \ket{\Downarrow} \, .
\end{equation}
Here $l=0, 1, \cdots, L$, $\ket{\Downarrow} = \otimes_{j} \ket{m_{j}=-1}$ is the fully polarized down state, $\hat{J}^{+} = \frac{1}{2} \sum_{j} (-1)^{j} (\hat{S}_{j}^{+})^{2}$  where $\hat{S}_{j}^{+} = \hat{S}_{j}^{x} + i \hat{S}_{j}^{y}$ is the ladder operator at site $j$. The normalization factor $\mathcal{N}_l$ is not essential for our discussion.
Under OBC, the system exhibits both spatial reflection symmetry and spin inversion symmetry in the zero total magnetization sector, and we shall perform computations in symmetry-resolved sectors.
The operator basis $\hat{L}_{j} = \left( \lambda_{a} \otimes \lambda_{b} \right)_{i,i+1}$ is chosen to be the same as that for the spin-1 AKLT model. Since the last term of the Hamiltonian \eqref{eq:Spin1_XY_OBC} does not belong to the operator space $\mathcal{V}$, the Hamiltonian will not manifest itself as a null vector of the correlation matrix.
As shown in Fig.~\ref{fig:Spin1_XY_OBC}, all thermal eigenstates have correlation-matrix zero $N_{0}=1$, which corresponds to the eigenoperator of total magnetization operator $\hat{M}_{z} = \sum_{i} \hat{S}_{i}^{z}$. In contrast, the scar state $\ket{\mathcal{S}_{L/2}}$ has a significantly larger $N_{0}=354$.

\begin{figure}[tbph]
    \centering
    \includegraphics[width=0.85\linewidth]{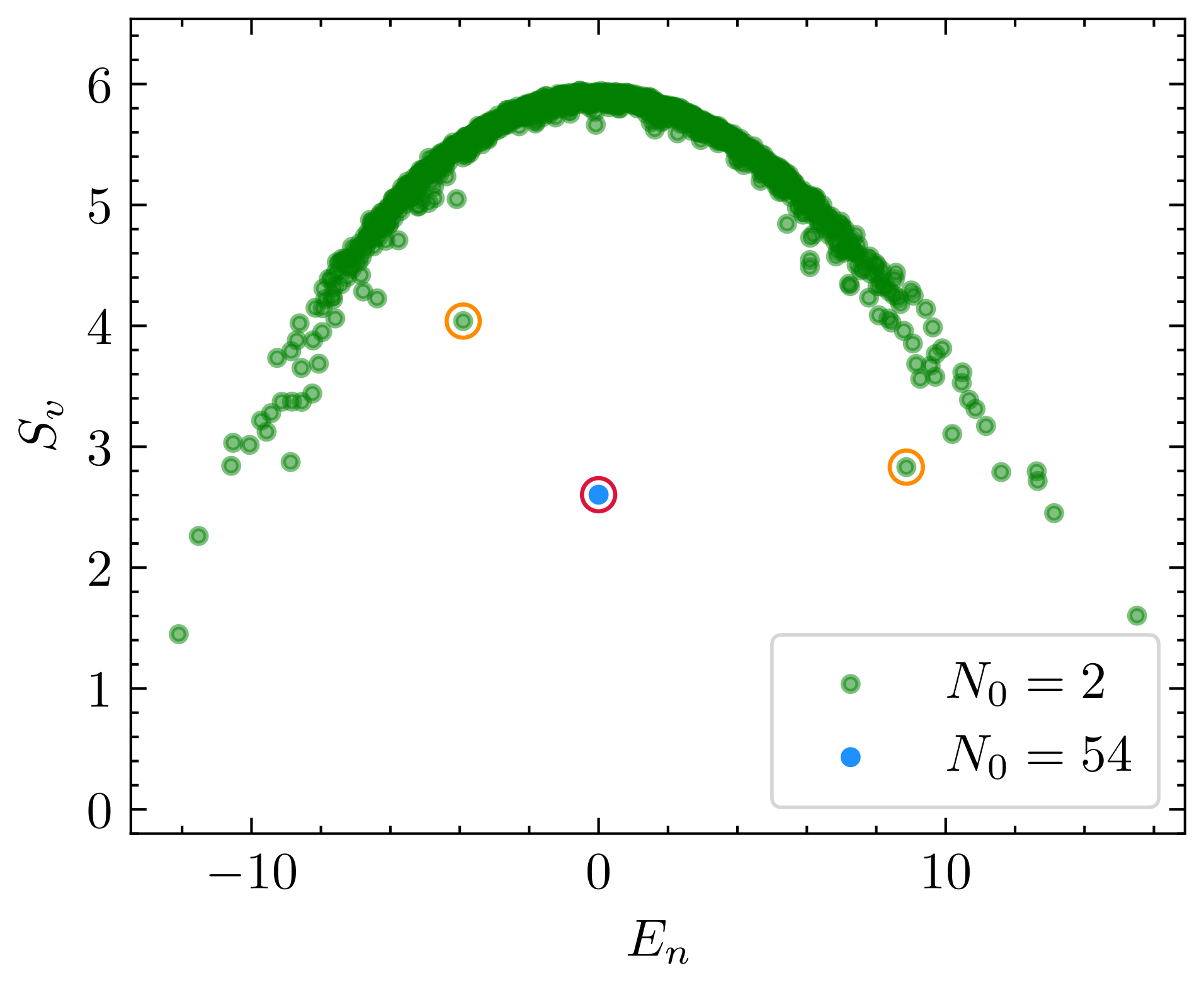}
    \caption{Plot of bipartite entanglement entropies of eigenstates against the eigenenergies $E_{n}$ for the spin-1 XY model \eqref{eq:Spin1_XY_PBC} with system size $L=12$, $g=0.2$, and $h=1.0$. The symmetry sector corresponds to total magnetization $M_{z}=0$, lattice momentum $k=0$, and both spatial reflection and spin inversion symmetric.}
    \label{fig:Spin1_XY_PBC}
\end{figure}

For periodic boundary conditions with $D=0$, the model admits another family of scar states \cite{Schecter_2019we}
\begin{align}
    \ket{\mathcal{S}_l^{\prime}} \propto\sum_{i_1 \neq i_2 \neq \cdots \neq i_l}  &  (-1)^{i_1+\cdots+i_l}  \left(S_{i_1}^{+} S_{i_1+1}^{+}\right) \nonumber \\
    & \left(S_{i_2}^{+} S_{i_2+1}^{+}\right) \cdots\left(S_{i_l}^{+} S_{i_l+1}^{+}\right) \ket{\Downarrow} \, .
\end{align}
However, due to a twisted SU(2) symmetry in even magnetization sectors, we study the following modified Hamiltonian
\begin{align}
    \hat{H} = & \sum_{\langle ij\rangle} \left(\hat{S}_i^x \hat{S}_j^x + \hat{S}_i^y \hat{S}_j^y \right) + h \sum_i \hat{S}_i^z \nonumber \\
    \label{eq:Spin1_XY_PBC}
    & + g \sum_{i} \left[ (\hat{S}_i^+)^2 (\hat{S}_{i+1}^-)^2 + \text{h.c.} \right] \, ,
\end{align}
where the last term serves to break integrability \cite{Chattopadhyay_2020qm}. This system preserves several symmetries: conservation of total magnetization $M_{z}$, lattice translation, spatial reflection, and spin inversion (which maps $\ket{m_{i}}$ to $\ket{-m_{i}}$ at site $i$). For our analysis, we focus on the $M_{z}=0$, lattice momentum $k=0$ sector that is symmetric under both spatial reflection and spin inversion.
The operator basis $\hat{L}_{j} = \left( \lambda_{a} \otimes \lambda_{b} \right)_{i,i+1}$ is chosen to be the same as the generalized AKLT model. As seen in Fig.~\ref{fig:Spin1_XY_PBC}, the thermal eigenstates uniformly have $N_{0}=2$, corresponding to the original Hamiltonian and the total magnetization operator $\hat{M}_{z} \equiv \sum_{i} \hat{S}_{i}^{z}$. The scar state $|\mathcal{S}'_{L/2}\rangle$ is distinguished by having a much larger $N_{0} = 54$.
Although this scar state can be identified by its low entanglement entropy, the correlation-matrix zero $N_0$ provides a \textit{qualitative} probe rather than a \textit{quantitative} one. Additionally, finite-size effects make it challenging to conclude whether the points circled in orange are scar states or not.

\begin{figure}[htbp]
    \centering
    \includegraphics[width=0.85\linewidth]{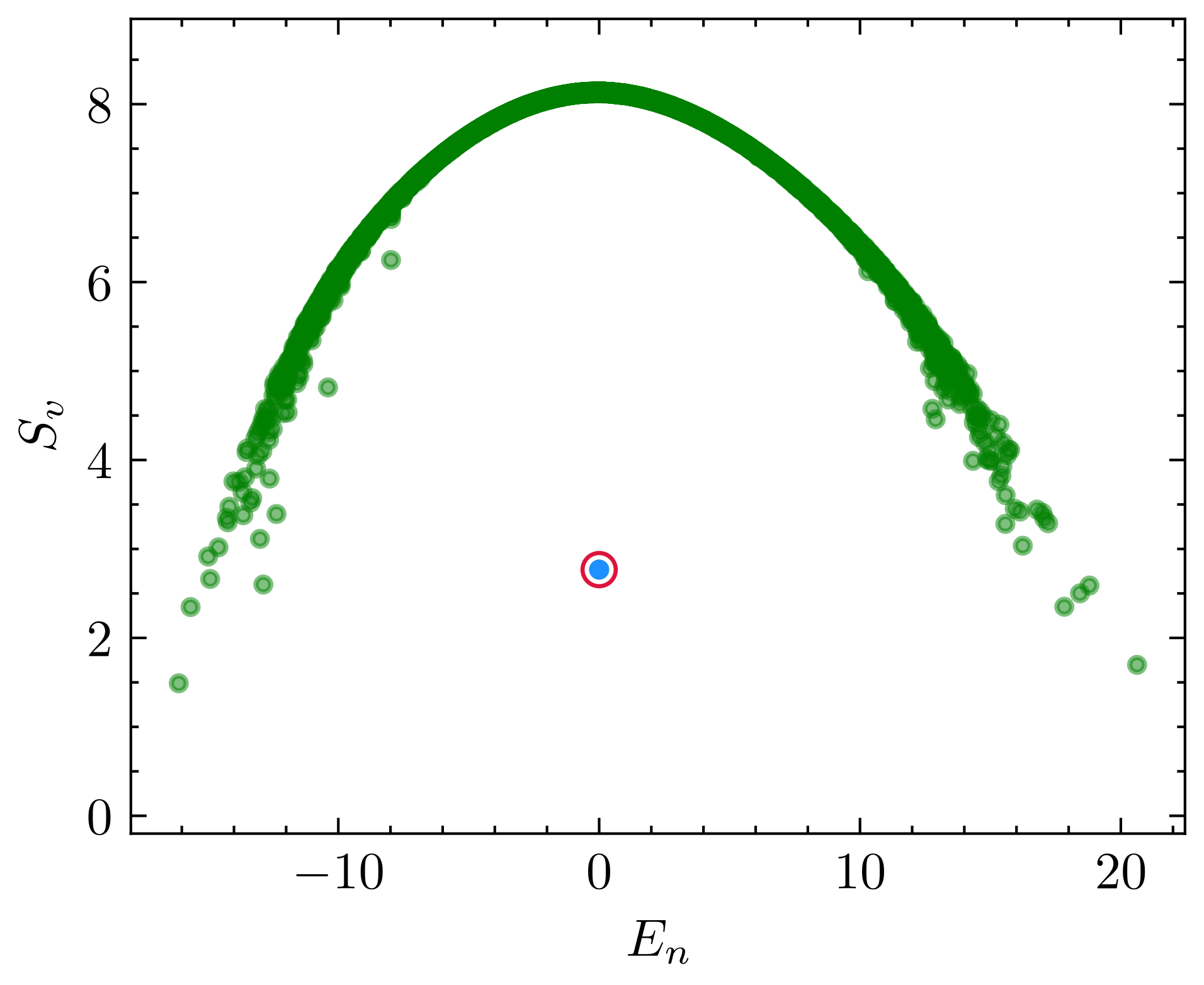}
    \caption{Plot of bipartite entanglement entropies of eigenstates against  eigenenergies $E_{n}$ for the spin-1 XY model \eqref{eq:Spin1_XY_PBC} with system size $L=16$. The parameters $g, h$ and the symmetry sector is the same as that in Fig.~\ref{fig:Spin1_XY_PBC}.}
    \label{fig:Spin1_XY_PBC_L16}
\end{figure}

After refining our algorithm, we successfully extended the entanglement entropy spectrum calculation to system size $L=16$. As can be seen from Fig.~\ref{fig:Spin1_XY_PBC_L16}, only the $E=0$ state (circled red) has anomalously low entanglement entropy compared to those of surrounding thermal eigenstates, which reinforces our confidence that the points circled in orange in Fig.~\ref{fig:Spin1_XY_PBC} are not QMBS.

\section*{APPENDIX D: APPROXIMATE EIGENOPERATORS IN THE PXP MODEL}
The PXP model distinguishes itself from scar-hosting models such as the generalized AKLT model and the spin-1 XY model in that the QMBS are approximate and escape an exact algebraic description in current formalisms. This is in line with our result that the QMBS share the same $N_{0}$, attributed to the Rydberg blockade condition, with the thermal eigenstates. However, as we show in Fig.~\ref{fig:PXP_Sv_Nc}(b), the correlation matrix of QMBS is marked by many approximate zero eigenvalues compared to that of the surrounding thermal eigenstates. We focus on the symmetry sector $(k, I)=(0, 1)$ for a system size $L=20$ as studied in Sec.~\ref{sec:approximate_QMBS}.
For definiteness, we consider the QMBS $\ket{\psi_{n}}$ with eigenenergy index $n=102$ and study its first approximate eigenoperator with the smallest nonzero eigenvalue $\lambda$ of the correlation matrix ($\lambda \approx 0.016$ compared to $\lambda \approx 0.1$ of the surrounding thermal eigenstates). The approximate eigenoperator $\hat{O}$ for $\ket{\psi_{n}}$ is numerically determined to be:
\begin{align*}
	\hat{O} = & \sum_{n} (-1)^{n} \Big[ c_{1} \left(\sigma_{z} \otimes I \otimes \sigma_{y} -  \sigma_{y} \otimes I \otimes \sigma_{z} \right)_{n,n+1,n+2} \\
	& + c_{2} \left( \sigma_{y} \otimes \sigma_{z} + \sigma_{z} \otimes \sigma_{y} \right)_{n,n+1}  \\
	& + c_{3} \left( \sigma_{y} \otimes \sigma_{x} \otimes \sigma_{x} - \sigma_{x} \otimes \sigma_{x} \otimes \sigma_{y} \right)_{n,n+1,n+2} \\
	& + c_{4} \left( \sigma_{x} \otimes \sigma_{y} + \sigma_{y} \otimes \sigma_{x} \right)_{n,n+1}  \\
	& + c_{5} \left( \sigma_{z} \otimes \sigma_{y} \otimes \sigma_{x} - \sigma_{x} \otimes \sigma_{y} \otimes \sigma_{z} \right)_{n,n+1,n+2} \\
	& + c_{6} \left( \sigma_{y} \otimes \sigma_{z} \otimes \sigma_{z} - \sigma_{z} \otimes \sigma_{z} \otimes \sigma_{y} \right)_{n,n+1,n+2}
	\Big] + \cdots
\end{align*}
where $c_{1} \approx  0.11304$, $c_{2} \approx  0.09789$, $c_{3} \approx 0.03133$, $c_{4} \approx 0.02641$, $c_{5} \approx 0.02498$, $c_{6} \approx 0.01515$ and contributions from much smaller $c_{i}$ have been omitted, the subscripts label the sites the operators act nontrivially. While the coefficients $c_{n}$ show no simple relation, the $(-1)^{n}$ factor is a robust feature for several other approximate eigenoperators we have numerically studied. \\

Motivated by the projection embedding approach to QMBS in the PXP model that first constructs an approximate wavefunction in the spin-1 basis and then applies the Rydberg projection operator $\hat{P}_{\text{Ryd}}$ to project out unphysical states \cite{Omiya_2023qm}, we then move to construct an approximate eigenoperator in the spin-1 basis. Due to the Rydberg blockade condition, two neighboring spins in the PXP model admit only three allowed configurations $\ket{00}$, $\ket{01}$, and $\ket{10}$. One can use a spin-1 basis $\ket{-}, \ket{0}$, and $\ket{+}$ to represent these three physical configurations: $\ket{10} = \ket{-}$, $\ket{00} = \ket{0}$, and $\ket{01} = \ket{+}$. The Rydberg projection operator in the spin-1 basis then becomes
\begin{equation}
	\hat{P}_{\text{Ryd}} = \prod_{n} \left( 1 - \ket{+-} \bra{+-} \right)_{n,n+1}
\end{equation}
where $n$ labels the spin-1 sites. For simplicity, we choose the Rydberg projected spin-1 operator $\hat{P}_{\mathrm{Ryd}} \hat{S}^{\alpha}$ with $\alpha = x,y,z$ to build a range-2 correlation matrix operator basis $\{\hat{L}_{i}\}$. We then follow the same steps as outlined in the Sec.~\ref{sec:CorrMat} to construct the correlation matrix for each eigenstate.
We find that all eigenstates have one \textit{exact} correlation-matrix zero, which turns out to be $\hat{O} = \hat{P}_{\mathrm{Ryd}} \sum_{n} \hat{S}_{n}^{x}$ where $n$ labels the $L/2$ sites in the spin-1 representation.
Since the PXP Hamiltonian simply flips spins subject to the Rydberg blockade condition, the exact eigenoperator $\hat{O}$ is nothing but the Hamiltonian in the spin-1 representation
\begin{equation}
	\hat{H} = \sqrt{2} \hat{P}_{\mathrm{Ryd}} \sum_{n} \hat{S}_{n}^{x} \, ,
\end{equation}
as the spin flip is fully accounted for by the $\hat{S}_{n}^{x}$ operator in the spin-1 representation. Therefore, the correlation matrix method also provides a tool to rewrite a Hamiltonian in a different representation. We then move on to study the first approximate eigenoperator of the same QMBS in the PXP model [$n=102$ in the symmetry sector $(k,I)=(0,1)$ for system size $L=20$]. The approximate eigenoperator $\hat{O}$ is numerically determined to be:
\begin{align*}
	\hat{O} & = \sum_{n} (-1)^{n} \Big[ c_{1} (S_{z} \otimes S_{z})_{n,n+1} + c_{2} (S_{x} \otimes S_{x})_{n,n+1} \\
	& + c_{3} \left(S_{y} \otimes S_{y} \right)_{n,n+1} \Big]  + \cdots
\end{align*}
where $(c_{1}, c_{2}, c_{3}) \approx (0.2194, 0.18615, 0.12957)$ and contributions from much smaller $c_{i}$ have been omitted. Still, the $(-1)^{n}$ factor is a robust feature.

%

\end{document}